\documentclass[a4paper,11pt]{article}
\usepackage{amssymb,amsmath,amsthm,amsfonts}
\usepackage{bookmark}
\usepackage{mathrsfs}
\usepackage{multirow}
\usepackage{graphicx}
\usepackage{authblk}
\usepackage{indentfirst}
\usepackage{multicol}
\usepackage{tabu}
\usepackage{tabularx}
\usepackage{url}
\usepackage{fancyhdr}
\usepackage[numbers]{natbib}
\usepackage[all]{xy}
\usepackage{mathtools}
\usepackage[english]{babel}
\usepackage{colortbl}
\usepackage{caption}
\usepackage{hyperref}
\usepackage{xcolor}
\usepackage[mathlines]{lineno}

\usepackage{xr}
\makeatletter
    \newcommand*{\addFileDependency}[1]{
    \typeout{(#1)}
    \@addtofilelist{#1}
    \IfFileExists{#1}{}{\typeout{No file #1.}}
    }
\makeatother

\title{Category-Specific Topological Learning of Metal-Organic Frameworks}

\author[1]{Dong Chen} 
\author[2]{Chun-Long Chen \thanks{Corresponding author: chunlong.chen@pnnl.gov}}
\author[1,3,4]{Guo-Wei Wei \thanks{Corresponding author: weig@msu.edu}}
\affil[1]{Department of Mathematics, Michigan State University, MI, 48824, USA}
\affil[2]{Physical Sciences Division, Pacific Northwest National Laboratory, Richland, Washington 99354, United States}
\affil[3]{Department of Electrical and Computer Engineering, Michigan State University, MI 48824, USA}
\affil[4]{Department of Biochemistry and Molecular Biology, Michigan State University, MI 48824, USA}

\makeatletter
    \renewcommand*{\@fnsymbol}[1]{\ensuremath{\ifcase#1\or \dagger\or *\or *\or
   \mathsection\or \else\@ctrerr\fi}}
\makeatother
\date{}

\begin{document}
    \maketitle

    \paragraph{Abstract}
    Metal-organic frameworks (MOFs) are porous, crystalline materials with high surface area, adjustable porosity, and structural tunability, making them ideal for diverse applications. However, traditional experimental and computational methods have limited scalability and interpretability, hindering effective exploration of MOF structure-property relationships. To address these challenges, we introduce, for the first time,  a category-specific topological learning (CSTL), which combines algebraic topology with chemical insights for robust property prediction. The model represents MOF structures as simplicial complexes and incorporates elemental categorizations to enable balanced, interpretable machine learning study. By integrating category-specific persistent homology, CSTL captures both global and local structural characteristics, rendering multi-dimensional, category-specific descriptors that support a predictive model with high accuracy and robustness across eight MOF datasets, outperforming all previous results. This alignment of topological and chemical features enhances the predictive power and interpretability of CSTL, advancing understanding of structure-property relationships of MOFs and promoting efficient material discovery.


    \newpage
    \tableofcontents
    \newpage

\section{Introduction}\label{section:introduction}

Metal-organic frameworks (MOFs) are a unique class of porous materials made up of metal ions or clusters connected to organic ligands, forming crystalline structures with remarkable tunability. Their customizable properties, including high surface area, adjustable porosity, and structural versatility, make MOFs highly suitable for a range of applications \cite{freund2021current,kumar2020green,chen2008guest,chen2005co2}, such as gas storage\cite{ma2010gas}, separation \cite{qian2020mof}, catalysis \cite{lee2009metal,kirlikovali2020zirconium}, and sensing \cite{kreno2012metal}. Although the design possibilities for MOFs are vast, with potentially infinite structures that can be synthesized. A thorough understanding of the relationship between MOF structure and its properties is therefore crucial for designing MOFs tailored to specific applications \cite{colon2014high,lee2021computational}. However, challenges remain. Traditional experimental methods, while valuable for providing insights into MOF behavior, can be labor-intensive, costly, and limited in scope, hindering the ability to explore the extensive chemical space that MOFs occupy. Computational methods, such as density functional theory (DFT) \cite{kohn1996density} and molecular dynamics (MD) \cite{karplus1990molecular}, enable detailed simulations of material behavior but often encounter scalability issues\cite{shao2022hierarchical}. These methods become computationally prohibitive, particularly for large or complex MOF systems, due to the extensive calculations required \cite{daglar2020recent,torkelson2024rational,zhao2023computational,yadav2024influence}.

Given the limitations of traditional experimental and computational approaches in studying MOF structure-property relationships, advanced data-driven techniques have become essential. Machine learning (ML) than has become increasingly important in studying MOF structure-property relationships and offering a possible solutions to those limitations \cite{wei2019machine,luo2022mof,chong2020applications,han2025development}. And thanks to the high-throughput computational screening, in particular, has emerged as a valuable approach, has laid a solid foundation by generating extensive, high-quality MOF databases\cite{banerjee2016metal,moghadam2024progress}, such as the CoRE MOF \cite{chung2019advances} and hMOF datasets \cite{wilmer2012large}, which enable ML applications in MOF research. Recently, ML models have leveraged geometric descriptors of MOF structures, such as void fraction and pore volume, to predict gas adsorption properties with notable accuracy \cite{orhan2021prediction,nandy2021using}. For instance, energy grid histograms have been used as descriptors in ML models to predict gas uptake \cite{bucior2019energy}, while other models utilize geometric, atom-type, and chemical feature descriptors to forecast N2/O2 selectivity and diffusivity \cite{orhan2021prediction}. Despite these advances, prediction accuracy remains a challenge for certain properties. The deep learning (DL) models are introduced, including convolution neural networks, graph neural networks \cite{rosen2021machine,xie2018crystal} and transformer-based architectures \cite{kang2023multi,park2023enhancing,cao2023moformer,chen2022interpretable}, have further enhanced the predictive power for various MOF properties by harnessing large datasets. However, these models come with certain limitations: they can be computationally demanding, often require substantial amounts of data, and sometimes function as 'black-box' systems, presenting challenges for interpretability. Addressing these considerations through continued refinement will help enhance the accessibility and interpretability of ML, particularly in advancing MOF discovery.


To address challenges in MOF research, incorporating mathematically derived, explainable features is essential. These features enhance interpretability and contribute to more robust predictive models for MOF properties. Instead of relying solely on conventional descriptors \cite{orhan2021prediction,nandy2021using}, advanced mathematical tools from fields like geometry and topology can be employed to extract insightful, high-level features. Techniques such as algebraic graph theory \cite{nguyen2019agl,chen2021algebraic}, persistent homology \cite{zomorodian2004computing}, element-specific persistent homology\cite{cang2018integration}, path topology \cite{chen2023path}, and topological Laplacians\cite{wang2020persistent} are increasingly used in molecular and materials science, offering new methods to capture the structural and functional nuances of complex materials. Mathematics-based methods have already shown success in fields such as drug discovery\cite{chen2021algebraic}, biological sciences \cite{cang2017topologynet}, and materials science \cite{jiang2021topological}, linking structural features to machine learning models for interpretable and detailed representations. For instance, persistent hyperdigraphs have enabled accurate predictions of protein-ligand interactions by capturing essential molecular details within a rigorous mathematical and transformer framework \cite{chen2024multiscale}. Mathematical deep learning was a top winner for pose and binding affinity prediction and ranking in D3R Grand Challenges, a worldwide competition series in computer-aided drug design \cite{nguyen2019mathematical,nguyen2020mathdl}. 

In this work, we propose a category-specific topological leaning (CSTL) model for predicting the properties of MOFs. This model introduces a mathematically sound and chemically informed framework designed to analyze and predict MOF properties by integrating both structural complexity and elemental composition. Specifically, each MOF structure is represented as a simplicial complex, establishing a robust topological basis for capturing the unique geometric features of MOFs. To enhance structural analysis with chemical insights, the model incorporates category-specific representations by categorizing elements based on valence electron similarity and occurrence frequency. This categorization ensures a balanced representation across the diverse elemental distributions of MOFs. For each elemental category, the model constructs tailored topological representations and applies persistent homology analysis. This method captures both global and local structural features using topological invariants, while also preserving detailed geometric information—particularly beneficial for materials with complex pore networks and spatially organized atomic structures. The model generates multi-dimensional, category-specific descriptors to encapsulate these intricate structural characteristics, which then serve as input to a gradient boosting tree model for predictive analysis. This approach provides an interpretable, chemically informed framework for predicting a broad range of MOF properties, including eight gas selectivity datasets, with the state-of-the-art performance and improved robustness. By aligning topological features with elemental distributions, CSTL addresses the limitations of conventional approaches, advancing the understanding and prediction of structure-property relationships in MOF materials.

\section{Results}

\subsection{Workflow and Schematic of a Category-Specific Topological Model}

\begin{figure}[!ht]
    \centering
    \includegraphics[width=16cm]{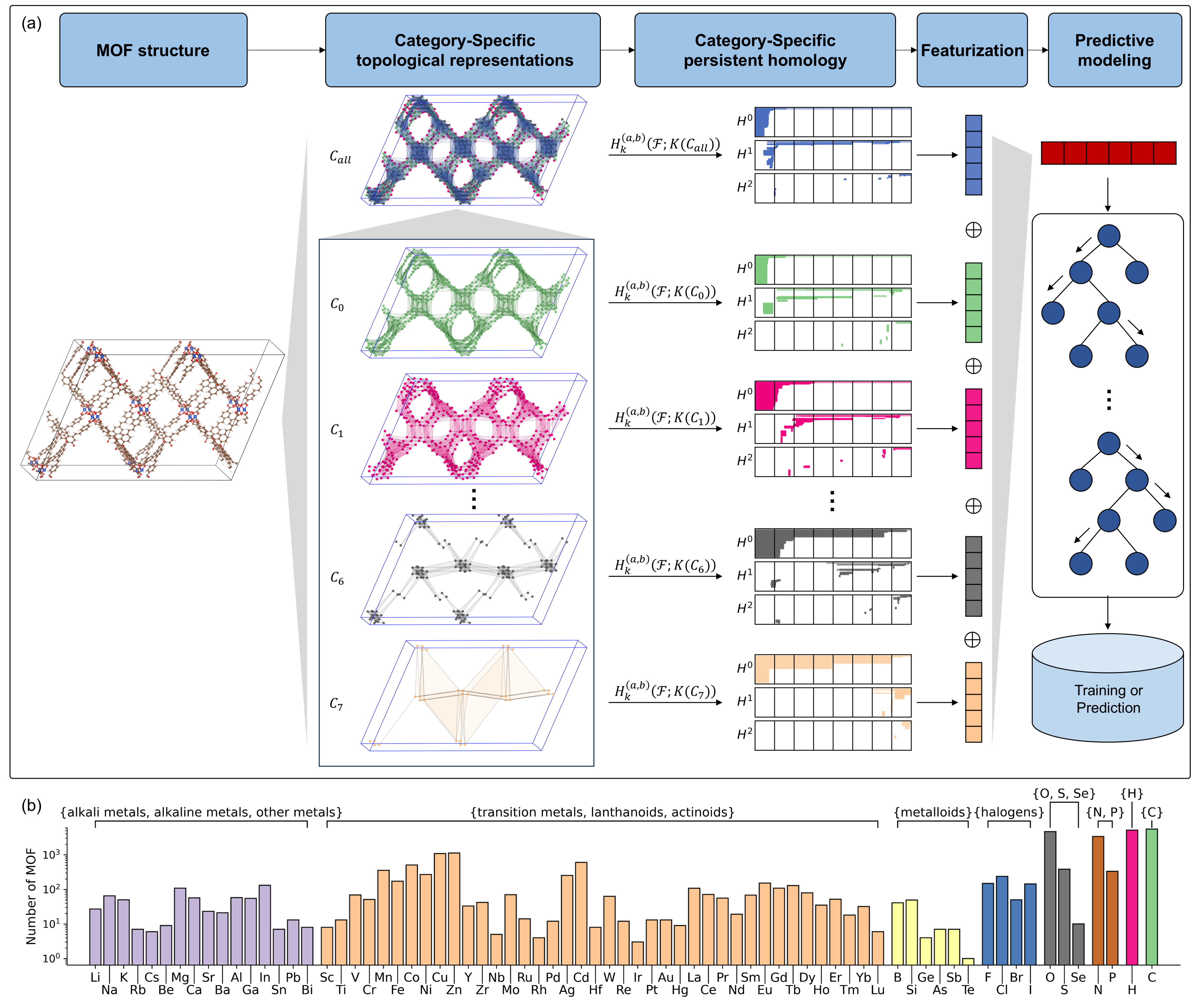}
    \caption{
        Schematic for Category-Specific Topological Models in MOF Property Prediction.  
        {\bf a} Overview of the category-specific topological models used for predicting properties of MOFs. Given MOF structures (first column), category-specific topological representations are constructed, including a simplicial complex for all structures ($C_{all}$) and sub-complexes based on categories ($C_0$ to $C_7$). The persistent homology method is applied to each category to generate barcode representations. A featurization vector extracts features from these barcodes, which are then used to construct a gradient boosting tree model for predictions on specific datasets.  
        {\bf b} Element distribution across the CoRE MOF v2019 dataset. The $y$-axis indicates the number of MOFs present in the dataset, with elements categorized based on valence electron similarity and their frequencies in the dataset.
    }
    \label{figure:workflow}
  \end{figure}

Figure~\ref{figure:workflow}{\bf a} presents the  workflow of our proposed category-specific topological model, designed to analyze and predict properties of MOF structures. In this workflow, the model begins by constructing a simplicial complex representation, which provides a robust topological framework tailored to capture the complex geometry of MOF materials. To enhance this analysis with chemical insights, we introduce category-specific topological representations, defined as categories $C_0$ through $C_7$ and $C_{all}$, where $C_{all}$ encompasses the full structure. As a preprocessing step, an elemental distribution analysis is conducted across the dataset, as illustrated in Figure~\ref{figure:workflow}{\bf b}, highlighting the frequency of elements in the dataset and grouping them by valence electron similarity and occurrence frequency, with distinct colors assigned to each group. Table~\ref{tbl:elementcategory} further specifies these element categories. Notably, the dataset exhibits a broad range of elements, with particular diversity among metallic elements, though specific metallic elements appear less frequently. The categorization process addresses these distribution variances, ensuring that infrequent elements are adequately considered within the predictive model to prevent overemphasis on particular elements' influence.

\begin{table}[h]
    \centering
    \caption{Element categories for Category-specific topological modeling of MOFs.}
    \label{tbl:elementcategory}
    \begin{tabular}{cc}
    \hline
    \textbf{Element Category} & \textbf{Notation} \\ \hline
    Alkali metals, alkaline metals, and other metals & $C_0$ \\ \hline
    Transition metals, lanthanoids, actinoids       & $C_1$ \\ \hline
    Metalloids                                      & $C_2$ \\ \hline
    Halogens                                        & $C_3$ \\ \hline
    Hydrogen (H)                                    & $C_4$ \\ \hline
    Carbon (C)                                      & $C_5$ \\ \hline
    Nitrogen (N), Phosphorus (P)                    & $C_6$ \\ \hline
    Oxygen (O), Sulfur (S), Selenium (Se)           & $C_7$ \\ \hline
    All                                             & $C_{all}$ \\ \hline
    \end{tabular}
    \end{table}

Based on these elemental categories, category-specific topological representations are constructed for each MOF structure, employing alpha complexes to provide a categorized-level topology for these materials. Subsequently, category-specific persistent homology analysis is applied, denoted as $H_{k}^{(a, b)}$, where $k = 0,1,2$ represents different topological dimensions, and $a = 0$ to $b = 25$ defines the distance interval, allowing a detailed examination of structure across multiple scales. Multi-dimensional category-specific barcodes are then computed to capture geometric and topological information specific to each elemental category. Following this, a featurization step bins these barcodes into intervals ranging from 0 to 25 \AA\ with a resolution of 0.1 \AA, producing length-fixed features from the barcodes. Finally, these features are concatenated to create a comprehensive and  category-specific topological descriptor, which is fed into a gradient boosting tree model for predictive modeling across various MOF properties. This approach ensures a balanced representation of elements within the model, enhancing predictive robustness and capturing the nuanced impacts of elemental distribution on MOF properties.

\subsection{Properties Prediction for MOFs}

\begin{figure}[!ht]
    \centering
    \includegraphics[width=16cm]{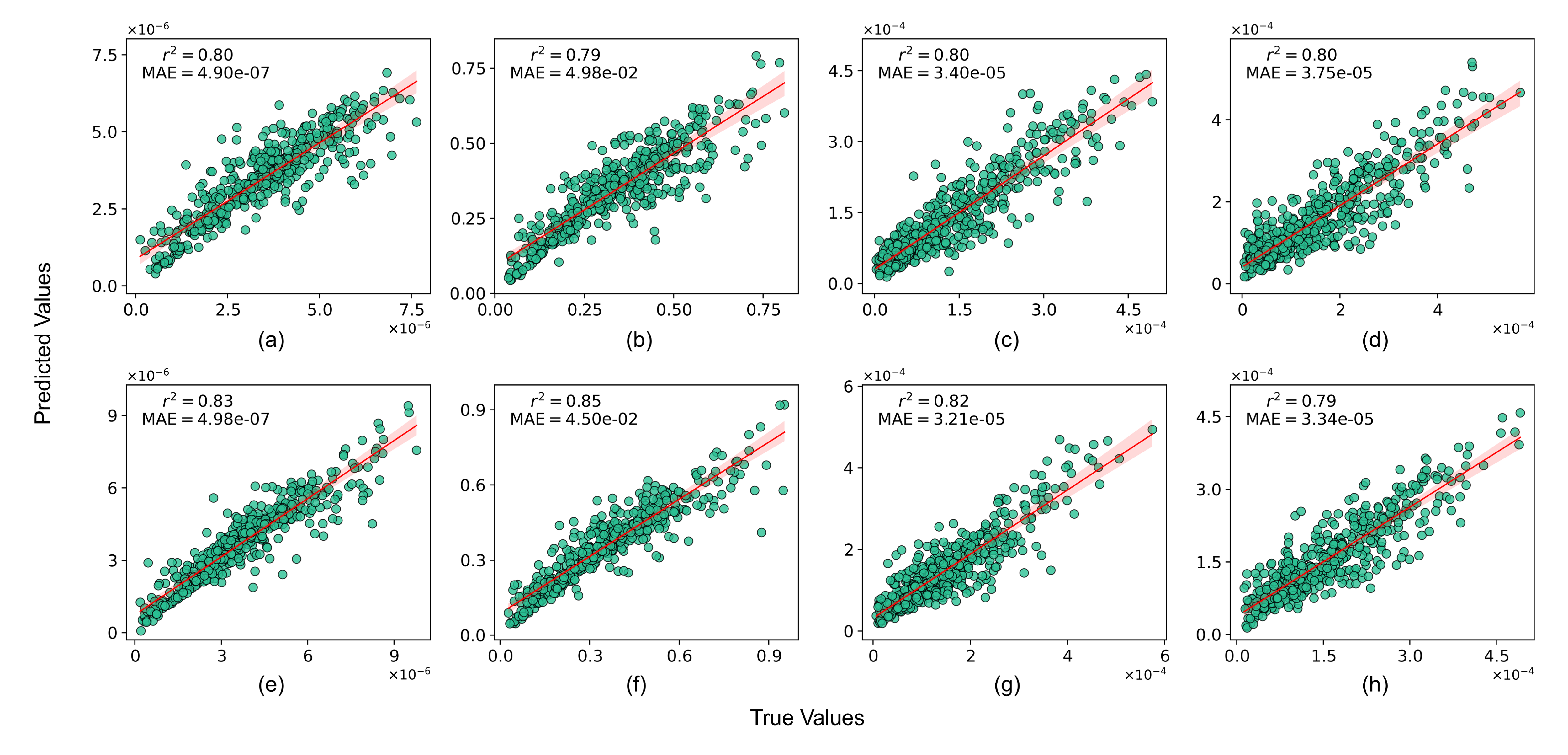}
    \caption{Comparison between predicted and true values for eight datasets on O$_2$/N$_2$ selectivity properties in MOF materials. Panels {\bf a}-{\bf h} show prediction performance for different properties: Henry's constant for N$_2$/O$_2$ (a, e), N$_2$/O$_2$ uptake (mol/kg) (b, f), self-diffusivity of N$_2$/O$_2$ at 1 bar (cm$^2$/s) (c, g), and self-diffusivity of N$_2$/O$_2$ at infinite dilution (cm$^2$/s) (d, h). Each panel displays the R$^2$ and the MAE in the upper left corner. Each dataset was randomly split, with 80\% used for training, with 10\% reserved for testing.}
    \label{figure:result}
  \end{figure}

    In this study, we validated the proposed category-specific topological models by predicting key properties of MOF materials, including the Henry's constant for N$_2$ and O$_2$ (mol/kg/Pa), uptake values for N$_2$ and O$_2$ (mol/kg), self-diffusivity of N$_2$ and O$_2$ at 1 bar (cm$^2$/s), and self-diffusivity at infinite dilution (cm$^2$/s). Table~\ref{tbl:dataset} and the Datasets section provide detailed descriptions of these datasets. The prediction outcomes, shown in Figure~\ref{figure:result}, demonstrate close alignment between predicted and actual values across all eight datasets, with an 80:10:10 random split for training, validation, and testing, respectively\cite{kang2023multi,park2023enhancing}. Performance metrics, specifically r$^2$ and MAE, averaged over 100 repeated experiments, are presented in the top left corner of each dataset's plot, underscoring the model's accuracy and reliability.

    To benchmark the model's performance, we compared it with state-of-the-art models, including MOFTransformer\cite{kang2023multi} and PMTransformer\cite{park2023enhancing}, both of which were trained on over a million structures for MOF property prediction. As shown in Table~\ref{tbl:comparison}, the category-specific topological model consistently outperforms these models across all datasets, achieving superior r$^2$, MAE, and RMSE metrics. It is noted that a universal set of hyperparameters was applied across all eight datasets to ensure robustness and prevent overfitting; validation data was not specifically used. In practical applications, incorporating the validation data into the training set could further enhance model accuracy.
    
\begin{table}[ht]
        \centering
        \caption{Comparison of CSTL performance with published models across various MOF datasets.}
        \label{tbl:comparison}
        \resizebox{1\textwidth}{!}{
        \begin{tabular}{ccccccccc}
        \hline
        \textbf{Datasets} & \multicolumn{3}{c}{\textbf{CSTL}} & \multicolumn{2}{c}{\textbf{Descriptor-based}\cite{orhan2021prediction}} & \multicolumn{2}{c}{\textbf{MOFTransformer}\cite{kang2023multi}} & \textbf{PMTransformer}\cite{park2023enhancing} \\
        \cline{2-9}
        & \textbf{r2} & \textbf{MAE} & \textbf{RMSE} & \textbf{r2} & \textbf{RMSE} & \textbf{r2} & \textbf{MAE} & \textbf{MAE} \\
        \hline
        Henry's constant N\textsubscript{2}                                                 & 0.80 & 4.90E-07 & 7.25E-07 & 0.70 & 8.94E-07 &      &          &          \\
        Henry's constant O\textsubscript{2}                                                 & 0.83 & 4.98E-07 & 7.63E-07 & 0.74 & 9.60E-07 &      &          &          \\
        N\textsubscript{2} uptake (mol/kg)                                                  & 0.79 & 4.98E-02 & 7.37E-02 & 0.71 & 8.62E-02 & 0.78 & 7.10E-02 & 6.90E-02 \\
        O\textsubscript{2} uptake (mol/kg)                                                  & 0.85 & 4.50E-02 & 6.82E-02 & 0.74 & 9.28E-02 & 0.83 & 5.10E-02 & 5.30E-02 \\
        Self-diffusion of N\textsubscript{2} at 1 bar (cm\textsuperscript{2}/s)             & 0.80 & 3.40E-05 & 4.69E-05 & 0.76 & 5.00E-05 & 0.77 & 4.52E-05 & 4.53E-05 \\
        Self-diffusion of N\textsubscript{2} at infinite dilution (cm\textsuperscript{2}/s) & 0.80 & 3.75E-05 & 5.15E-05 & 0.76 & 5.50E-05 &      &          &          \\
        Self-diffusion of O\textsubscript{2} at 1 bar (cm\textsuperscript{2}/s)             & 0.82 & 3.21E-05 & 4.45E-05 & 0.78 & 4.98E-05 & 0.78 & 4.04E-05 & 3.99E-05 \\
        Self-diffusion of O\textsubscript{2} at infinite dilution (cm\textsuperscript{2}/s) & 0.79 & 3.34E-05 & 4.53E-05 & 0.74 & 4.95E-05 &      &          &          \\
        \hline
        \end{tabular}}
    \end{table}

    Additionally, we evaluated model robustness by testing on a 20\% holdout set across all datasets, with results shown in Figure S1 and Table S1, where the proposed model continued to outperform previous models. To ensure the validation stability, we trained 100 models using 10 different seeds, each repeated across 10 randomly initialized predictive models. Heatmaps in Figures S2, S3, and S4 illustrate that variations in seed selection have minimal impact on model performance, confirming the robustness and stability of the predictive model across both fixed and variable data splits.


\subsection{Feature analysis for the category-specific topological models}

    In this work, we propose a category-specific topological model to capture the distinctive characteristics of MOF materials. The model encodes each component's inherent structural and functional attributes by categorizing elements based on their chemical roles and applying persistent homology to analyze each category separately. This approach allows us to represent both the inorganic and organic building blocks of MOFs through distinct topological features, providing a nuanced view that goes beyond treating all atoms as identical.

    MOFs are typically built from two primary types of components: inorganic metal nodes and organic linkers. Metal ions or clusters in the inorganic units serve as coordination centers and framework backbones, offering stability and structural rigidity while connecting to the organic linkers. Although metal nodes often appear in smaller quantities than organic atoms, they strongly influence the overall material properties\cite{chen2005co2,rohde2024high}. Because of the diversity among metal elements, it becomes challenging to systematically understand the effect of each metal across all samples—especially for rare metals like Rn, Bi, and Cs that appear infrequently. Organic linkers, composed mainly of carboxylates or nitrogen-containing ligands, bridge these metal nodes, defining the MOF's porosity and connectivity. These organic components typically make up the majority of the framework and play a critical role in establishing the intricate, symmetrical structures of MOFs.

    To address these component-specific influences, we group metals into categorical types ($C_0$, $C_1$, $C_2$, $C_3$) while non-metals are clustered into single element or few elements set ($C_4$, $C_5$, $C_6$, and $C_7$) as shown in Table~\ref{tbl:elementcategory}. This CSTL thus captures the functional contributions of distinct components within the MOF without overemphasizing elemental diversity, allowing each category to reveal its unique structural influence through topological embedding.

\begin{figure}[!ht]
    \centering
    \includegraphics[width=12cm]{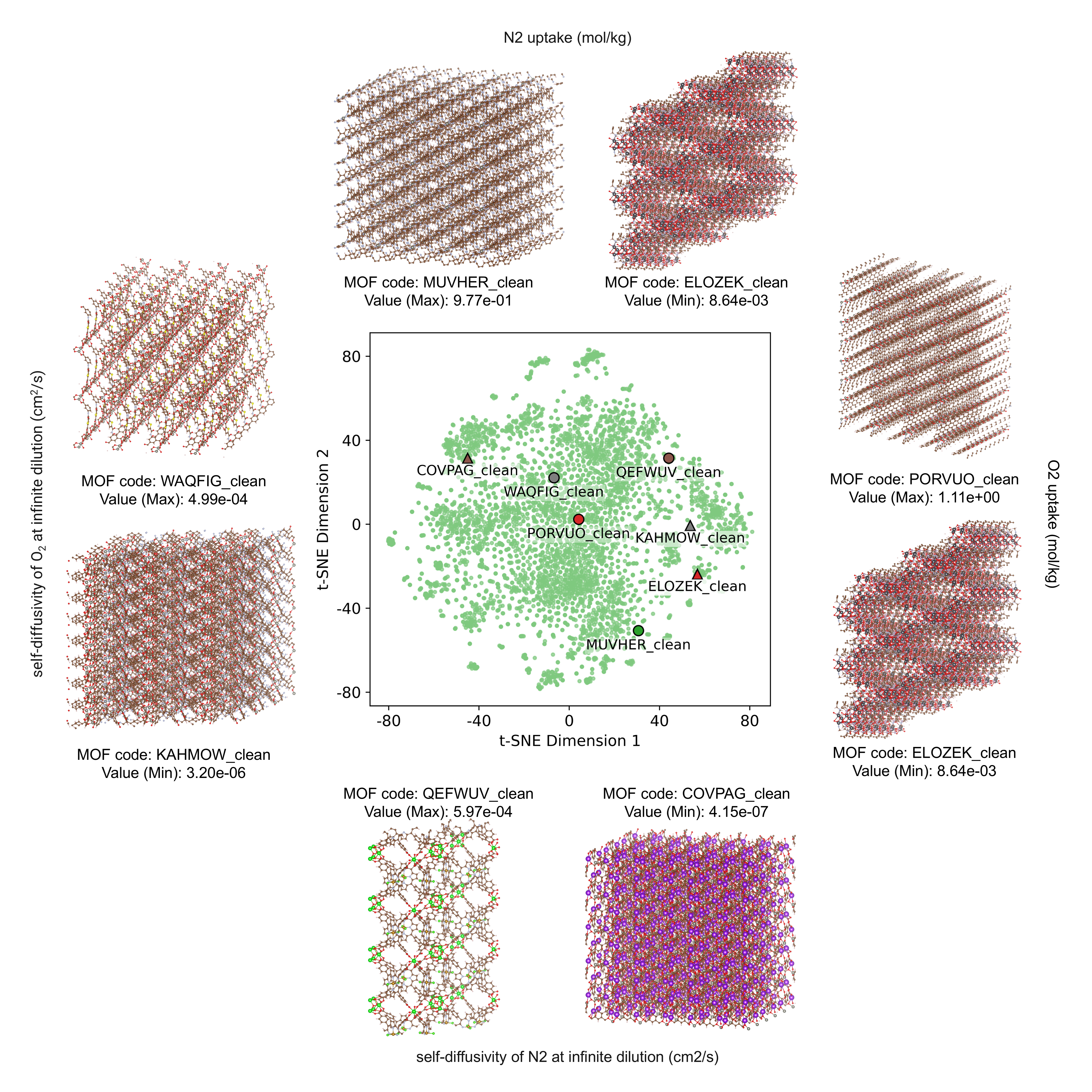}
    \caption{t-SNE feature reduction for category-specific topological features of MOF materials, where each green point represents a distinct MOF material. Highlighted circles and triangles indicate materials with maximum and minimum values, respectively, for four key properties: N$_2$ uptake (mol/kg), O$_2$ uptake (mol/kg), self-diffusivity of N$_2$ at infinite dilution(cm$^2$/s), and self-diffusivity of O$_2$ at infinite dilution(cm$^2$/s). 3D structures of the materials with minimum and maximum values for each property are shown around the t-SNE plot.}
    \label{figure:embedding}
  \end{figure}

    Visualizing the 2D t-SNE reduction in Figure~\ref{figure:embedding}, each green point represents a different MOF material, with distinct clusters reflecting the influence of the CSTL features. Here, key properties such as N$_2$ uptake, O$_2$ uptake, and self-diffusivity values are mapped, where materials with the maximum and minimum values for each property are highlighted. Even without predictive modeling, CSTL features differentiate structures with significant property variations, suggesting that the model inherently captures critical structure-property relationships. For example, the MOF material labeled ELOZEK\_clean, which has the lowest N$_2$/O$_2$ uptake values (8.64e-03 mol/kg for both N$_2$ and O$_2$) and Henry's constants (8.64e-03 mol/kg/Pa for both N$_2$ and O$_2$), reflects poor gas absorption. Similarly, COVPAG\_clean demonstrates minimal self-diffusivity for N$_2$ (4.15e-07 cm$^2$/s), underscoring its limited diffusion capabilities. Such distinctions underscore the power of the CSTL approach to reveal essential structural variations directly through category-specific topological embeddings, distinguishing materials with extreme property values across the MOF dataset.

\begin{figure}[!ht]
    \centering
    \includegraphics[width=16cm]{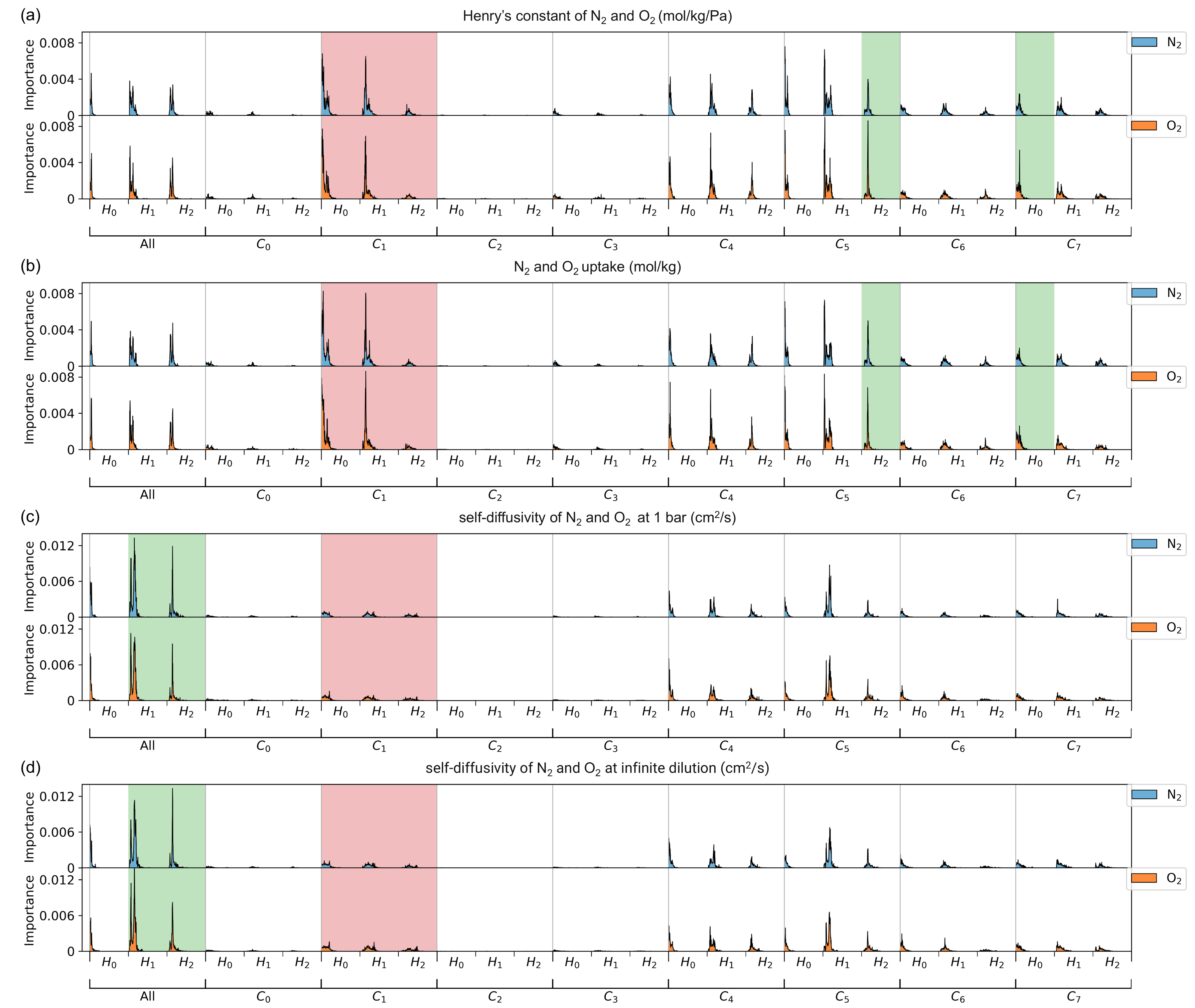}
    \caption{Feature importance analysis for predictive models of eight properties in MOF materials using gradient boosting tree model-based importance. Panels (a)-(d) show the importance of topological features for predicting (a) Henry's constant of N$_2$ and O$_2$ (mol/kg/Pa), (b) N$_2$ and O$_2$ uptake (mol/kg), (c) self-diffusivity of N$_2$ and O$_2$ at 1 bar (cm$^2$/s), and (d) self-diffusivity of N$_2$ and O$_2$ at infinite dilution (cm$^2$/s). Each panel presents separate importance values for N$_2$ (blue) and O$_2$ (orange) predictions. The bars highlight feature groups by topological dimensions $H_0 $, $H_1 $, and $H_2 $ across different topological categories ($C_0 $ to $C_7 $), with green-shaded regions indicating particularly influential features for different gases (N$_2$ and O$_2$) and red-shaded regions indicating particularly influential features across different properties.}
    \label{figure:featureimportance}
  \end{figure}

    To quantify the significance of each feature within the proposed CSTL model, we analyzed the tree-based feature importance derived from trained predictive models, as illustrated in Figure~\ref{figure:featureimportance}. This analysis highlights several key trends across different homology dimensions ($H_0$, $H_1$, and $H_2$) and categories, reflecting the structural and categorical influence on the model's predictions.

    Generally, we observe that feature importance is concentrated at the beginning of each dimensional homology ($H_0$, $H_1$, and $H_2$) across all categories. This is due to the intentionally large end value set for the intervals (25 \AA\ ), ensuring the model's robustness across a broader range of structures, including potential extreme cases beyond the current dataset. Consequently, topological features in the later portion of the interval largely default to zero, explaini ng the higher importance of features at the beginning of each homology dimension. For category $C_2$, which includes metalloids like B, Si, Ge, As, Sb, Te, Po, and At, the feature importance appears limited. Since these elements have valence electron configurations similar to carbon, and their occurrence within the dataset is low (as shown in Figure~\ref{figure:workflow}{\bf b}), their influence is often overshadowed by the predominant presence of carbon. This results in carbon having a stronger impact within this category, affecting the overall model importance distribution.

    Focusing on Henry's constant, shown in the green-highlighted section of Figure~\ref{figure:featureimportance}{\bf a}, we see distinct variations in feature importance between different gases (N$_2$ in blue and O$_2$ in orange). Categories $C_5$ and $C_7$, representing carbon and oxides respectively, exhibit substantial shifts in importance, indicating that carbon-based structures and strongly oxidizing elements influence the selectivity of MOF materials towards these gases. In particular, $H_2$ in $C_5$ suggests that carbon-based cavities strongly affect gas selectivity, while $H_0$ in $C_7$ highlights the role of oxidizing element spacing on selectivity. A similar trend is observed for N$_2$ and O$_2$ uptake properties, as shown in Figure~\ref{figure:featureimportance}{\bf b}. For self-diffusivity of N$_2$/O$_2$, whether at 1 bar or infinite dilution, Figure~\ref{figure:featureimportance}{\bf c} and {\bf d} indicate that cycles and cavities within the overall MOF structure, particularly within $H_1$ and $H_2$ of the $C_{all}$ category, are the primary factors influencing diffusion properties. This suggests that the model effectively captures the topological elements critical to gas diffusion across  MOF structures.

    Furthermore, when comparing properties related to gas absorption (Figure~\ref{figure:featureimportance}{\bf a} and {\bf b}) and diffusivity (Figure~\ref{figure:featureimportance}{\bf c} and {\bf d}), we note that $C_1$ shows significant variations in importance. This implies that metal atoms have a pronounced effect on gas absorptivity, in contrast to their relatively lower impact on diffusivity properties. In conclusion, this feature analysis demonstrates the versatility and precision of the proposed CSTL model, which adeptly balances generalization and prediction accuracy across diverse property predictions. By integrating both structural and elemental distinctions, the model captures the nuanced interactions within MOF materials, offering a robust framework for predicting many functional properties.

\section{Method}
\subsection{Datasets} \label{subsection:dataset}

    All datasets used in this study originate from the CoRE MOFs 2019 database \cite{chung2019advances}. The properties of interest, such as O$_2$ and N$_2$ selectivity, were simulated in earlier studies \cite{chung2019advances}. Certain entries, identified as outliers, were found to lie at the extreme upper end of the distribution, significantly distant from the majority of data points \cite{orhan2021prediction}. To enhance the robustness and reliability of the models, these outliers were excluded. Additionally, the data used for validating property predictions was further refined by applying upper-limit threshold values, as outlined in the work of \citet{orhan2021prediction}. This filtering process removed the outliers, resulting in a more uniform and comprehensive distribution, ensuring a well-represented target-variable space. The filtering methodology follows that of \citet{orhan2021prediction}. To ensure a clear and transparent comparison, we compiled the data information for all methods compared in this study, as summarized in Table S1
    In this study, the input features were derived solely from MOF structural data stored in CIF files, without the use of any additional descriptors. Detailed information about  the datasets, including the filtered properties and prediction performance, is provided in Table \ref{tbl:dataset}.

    \begin{table}[htbp]
        \centering
        \caption{Summary of datasets for N$_2$ and O$_2$ selectivity of MOFs.}
        \label{tbl:dataset}
        \resizebox{1\textwidth}{!}{
        \begin{tabular}{lccc}
        \hline
        \textbf{Datasets (Properties)} & \textbf{Sizes} & \textbf{Train:Valide:Test} & \textbf{Splitting Method} \\ \hline
        Henry's constant of N$_2$ (mol/kg/Pa) & 4744 & 80:10:10 & Random split \\
        Henry's constant of O$_2$ (mol/kg/Pa) & 5036 & 80:10:10 & Random split \\
        N$_2$ uptake (mol/kg)                 & 5132 & 80:10:10 & Random split \\
        O$_2$ uptake (mol/kg)                 & 5241 & 80:10:10 & Random split \\
        self-diffusivity of N$_2$ at 1 bar (cm2/s) & 5056 & 80:10:10 & Random split  \\ 
        self-diffusivity of N$_2$ at infinite dilution (cm2/s) & 5192 & 80:10:10 & Random split  \\
        self-diffusivity of O$_2$ at 1 bar (cm2/s) & 5223 & 80:10:10 & Random split  \\
        self-diffusivity of O$_2$ at infinite dilution (cm2/s) & 5097 & 80:10:10 & Random split  \\ \hline
        \end{tabular}}
    \end{table}

\subsection{Category-Specific Topology}


    \paragraph{Simplicial complex representations}

\begin{figure}[!ht]
        \centering
        \includegraphics[width=13.5cm]{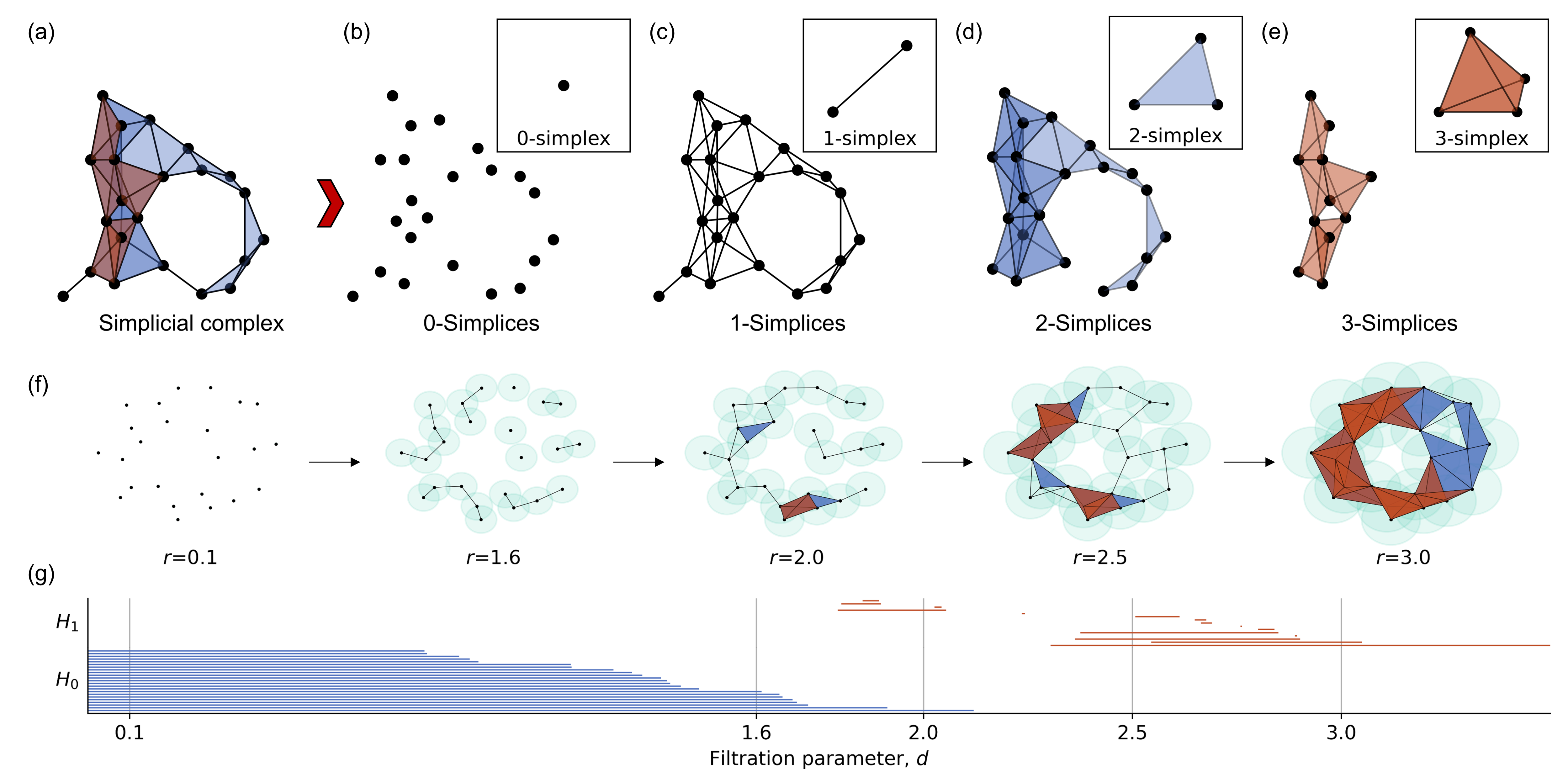}
        \caption{Illustration of concepts in persistent homology. (a) An example of a simplicial complex. (b)-(e) Expansion of the simplicial complex in (a) into different simplex dimensions: (b) 0-simplices (vertices), with the 0-simplex as a building block shown in the upper right; (c) 1-simplices (edges), with the 1-simplex in the upper right; (d) 2-simplices (triangles), with the 2-simplex in the upper right; and (e) 2-simplices, with the 3-simplex (tetrahedron) in the upper right. (f) A nested simplicial complex with an increasing parameter $r$, exemplifying an alpha complex. (g) Barcode representation of $H_0$ and $H_1$ for the complex in (f), with specific values of $r = 0.1$, 1.6, 2.0, 2.5, and 3.0 corresponding to the simplicial complex states shown in (f).}
        \label{figure:method}
    \end{figure}

    Simplicial complexes extend the concept of graphs to higher dimensions, offering richer structural and topological information, as shown an example in Figure~\ref{figure:method}{\bf a}. A $k$-simplex, defined as the convex hull of $k+1$ independent points, generalizes concepts like points (0-simplex, Figure~\ref{figure:method}{\bf b}), line segments (1-simplex, Figure~\ref{figure:method}{\bf c}), triangles (2-simplex, Figure~\ref{figure:method}{\bf d}), and tetrahedra (3-simplex, Figure~\ref{figure:method}{\bf e}). A $k$-simplex is the $k$-dimensional analog of these shapes, defined as the convex hull of $k+1$ affinely independent points, and can be expressed as
    \begin{equation}
        \sigma^{k} = \Big \{ v \, \big| \, v = \sum_{i=0}^{k} \lambda_{i} v_{i}, \sum_{i=0}^{k} \lambda_{i} = 1, \, 0 \leq \lambda_{i} \leq 1, \, i = 0, 1, \dots, k \Big \}.
    \end{equation}
    A simplicial complex $K$ is a collection of simplices such that (1) every face of a simplex in $K$ is also in $K$, and (2) the intersection of any two simplices is either empty or a common face.
    
    In this work, we represent MOF structures using simplicial complexes, where atoms are 0-simplices (vertices), bonds are 1-simplices (edges), and higher-order interactions, such as atomic rings and cavities, are captured as higher-dimensional simplices. This approach allows us to model not only the pairwise connections but also the higher-order geometric and topological features essential for understanding the physical and chemical properties of MOFs. In the category-specific representation framework for MOF structures, all atoms are grouped into distinct sets based on the categories listed in Table 1, denoted as $C_0$ to $C_7$. Additionally, $C_{all}$ represents the set containing all atoms. For each category-specific set, topological representations are constructed to capture the interactions among atoms across different categories.
    
    \paragraph{Homology and persistent homology}
    To analyze the topological properties of MOF structures represented by simplicial complexes, homology is used as an algebraic tool. By employing concepts such as chains, chain groups, and boundary operators, homology groups capture features like connected components, loops, and cavities within the material. For a given dimension $k$, the $k$-chain group $C_k$ is formed by $k$-simplices with coefficients from a specified field (e.g., $\mathbb{Z}_2$). The boundary operator $\partial_k$ maps $k$-chains to $(k-1)$-chains, defined as:
    \begin{equation}
      \partial_k [v_0, v_1, \ldots, v_k] = \sum_{i=0}^{k} (-1)^i [v_0, \ldots, \hat{v}_i, \ldots, v_k],
    \end{equation}
    where $\hat{v}_i$ indicates the omission of vertex $v_i$. This operation helps identify cycles (chains with no boundary) and boundaries (chains that are boundaries of higher-dimensional simplices). The $k$-th homology group $H_k$ is then defined as:
    \begin{equation}
    H_k = \text{Ker}(\partial_k) / \text{Im}(\partial_{k+1}),
    \end{equation}
    which represents $k$-dimensional holes, such as connected components ($H_0$), loops ($H_1$), and voids ($H_2$) in the MOF structure. The homology group ($H$) allows for the measurement of topological features, such as Betti numbers ($\beta$), which count the number of independent $k$-dimensional cycles, reflecting the number of $k$-dimensional holes. They can be calculated by, $\beta_k = \text{rank}(H_k) = \text{rank}(Z_k) - \text{rank}(B_k)$.

    To capture how these topological features vary with the spatial scale, persistent homology is introduced \cite{edelsbrunner2001geometry,zomorodian2005topology,zomorodian2004computing}. It tracks the evolution of homological features as a parameter (e.g., bond length or distance threshold) changes. This is achieved through a filtration, a sequence of nested subcomplexes $\{ K_i \}$ where $K_0 \subseteq K_1 \subseteq \ldots \subseteq K_n$. There are some common used filtration methods, such Vietoris-Rips comple \cite{hausmann1994vietoris}, Cech complex \cite{ghrist2014elementary}, and alpha complex \cite{edelsbrunner1995smooth}. In this work, we employ the alpha complex for analyzing MOF structures. The alpha complex is constructed based on the Delaunay triangulation of the atomic positions. For a given parameter $\alpha$, a simplex (e.g., an edge, triangle, or tetrahedron) is included in the complex if the radius of the smallest empty circumsphere that encloses it is less than or equal to $\alpha$. As $\alpha$ increases, the alpha complex grows, progressively capturing larger topological features in the MOF structure, such as rings, tunnels, and cavities, an example is shown Figure~\ref{figure:method}{\bf f}.
    
    Persistent homology quantifies the persistence of these features across different scales, revealing stable patterns that correspond to critical geometric and chemical properties of the MOF. Each $k$-th homology group is tracked across the filtration, providing insights into how certain features (e.g., porosity or connectivity) appear, merge, and disappear as the structure evolves. These persistent patterns are typically visualized using barcodes \cite{ghrist2008barcodes}, where the length of each bar represents the lifespan of a particular topological feature. An example of barcodes with corresponding alpha complex is shown in Figure~\ref{figure:method}{\bf g}.
        
    \paragraph{Category-Specific topological embedding}
    
    Persistent homology does not distinguish different element types and thus gives a poor representation for chemical and biological systems. Element-specific persistent homology was introduced to better capture chemical and biological properties \cite{cang2018integration}. 
    In this work, we propose a category-specific topological embedding approach to preserve the chemical and physical information inherent in MOF structures. Elements in the periodic table are categorized into eight distinct groups based on their chemical similarities and structural roles (see Table~\ref{tbl:elementcategory}). Before constructing the embeddings, the supercell of each material is scaled uniformly to approximately 64 \AA\ $\times$ 64 \AA\ $\times$ 64 \AA\ to ensure that the topological analysis is performed consistently across different structures.

    Our method involves two main stages: (1) For a given MOF structure, category-specific topological representations are constructed based on the elemental types of atoms, categorized as $C_0$ to $C_7$, along with an additional set $C_{all}$ containing all atoms. (2) The persistent homology of each category from stage (1) is computed to capture global and category-level topological patterns, characterized by their Betti numbers in the $H_0$, $H_1$, and $H_2$ homology spaces. This approach allows the topological analysis to incorporate both structural and chemical information. For each category and each homology dimension, we employ a grid-based method to generate the topological embeddings. Specifically, we construct a grid ranging from 0 to 25 \AA\, with a step size of 0.1 \AA\, and record the Betti numbers (i.e., the number of topological features that persist at each scale). This process yields a feature vector of length 750 (250 steps $\times$ 3 homology dimensions: $H_0$, $H_1$, and $H_2$) for each element category. By concatenating these feature vectors across all eight categories, we obtain a 6000-dimensional representation. When combined with the features derived from the entire MOF structure, the final topological embedding results in a 6750-dimensional vector that integrates both global structural patterns and category-specific chemical information.

\subsection{Predictive Modeling}
    In this work, a Gradient Boosting Tree (GBT) model was constructed to perform regression analysis using the proposed category-specific topological embedding as input features. Gradient boosting is an ensemble learning method that builds multiple weak learners (typically decision trees) sequentially, where each tree is trained to correct the errors made by the previous ones, thereby producing a more accurate model. We implemented the gradient boosting regressor from Scikit-learn \cite{pedregosa2011scikit}, optimizing the squared error loss function. The model parameters were set as follows: \text{max\_depth=7}, \text{max\_features=`sqrt'}, \text{min\_samples\_leaf=1}, \text{min\_samples\_split=2}, \text{n\_estimators=10,000}, and \text{subsample=0.5}. These settings were not fine-tuned, as we aimed to demonstrate the robustness of the proposed predictive model with a single set of hyperparameters.

    All input features were normalized using standard scaling, and the target properties were standardized to facilitate regression analysis. For model evaluation, we split the dataset into train, validation, and test sets using an 80\%, 10\%, and 10\% ratio, respectively \cite{kang2023multi,orhan2021prediction}. Since we used a universal set of hyperparameters, the validation set was not employed for model selection. Instead, 80\% of the data was used for training to establish a fair comparison with previous works. The results for the test set (10\%) and for both the test and validation sets combined (20\%) are reported to assess the model's performance comprehensively.

    To ensure robust evaluation, we repeated the random data split 10 times, and for each split, 10 models were trained with different random seeds, resulting in a total of 100 models per dataset. The performance metrics, including root mean square error (RMSE), mean absolute error (MAE), and $r^2$ correlation, were averaged over these 100 models and reported as the final results (as seen in Supplementary Information Section 1). 
     This approach of using a single set of hyperparameters and a consistent evaluation protocol highlights the robustness of the predictive model, making the results reliable and comparable to existing methods in the literature.

\section*{Data availability}
The datasets utilized in this work are derived from the structures available in the CoRE MOFs 2019 database \cite{chung2019advances}. The properties for each dataset were obtained using the methods outlined in \citet{orhan2021prediction}.


\section*{Acknowledgments}
This work was supported in part by NIH grants R01GM126189, R01AI164266, and R35GM148196, National Science Foundation grants DMS2052983 and IIS-1900473, Michigan State University Research Foundation, and  Bristol-Myers Squibb 65109.
C.-L.C. gratefully acknowledges financial support from the Defense Threat Reduction Agency (Project CB11141), and the Department of Energy (DOE), Office of Science, Office of Basic Energy Sciences (BES) under an award FWP 80124 at Pacific Northwest National Laboratory (PNNL). PNNL is a multiprogram national laboratory operated for the Department of Energy by Battelle under Contract DE-AC05-76RL01830.

\newpage
\appendix
\section{Evaluation metrics}\label{ssection:evaluation_metrices}

In this work, the Root Mean Square Error (RMSE), Mean Absolute Error (MAE), and $ R^2 $ (Coefficient of Determination) were used for evaluating machine learning models.

The Root Mean Square Error (RMSE) is a standard way to measure the error of a model in predicting quantitative data. It quantifies the difference between the predicted values ($ \hat{y}_i $) and the true values ($ y_i $).  The formula for RMSE is defined as:

\begin{equation}
\text{RMSE} = \sqrt{\frac{1}{n} \sum_{i=1}^{n} (\hat{y}_i - y_i)^2}
\end{equation}

where:
1) $ n $ is the number of observations.
2) $ \hat{y}_i $ is the predicted value for the $ i $-th observation.
3) $ y_i $ is the actual value for the $ i $-th observation.
4) Lower values of RMSE indicate a better fit of the model to the data. It has the same unit as the response variable.

The Mean Absolute Error (MAE) measures the average magnitude of errors in a set of predictions, without considering their direction. It calculates the average of the absolute differences between predicted and actual values. The formula for MAE is given by:

\begin{equation}
\text{MAE} = \frac{1}{n} \sum_{i=1}^{n} |\hat{y}_i - y_i|
\end{equation}

where:
1) $ n $ is the number of observations.
2) $ \hat{y}_i $ is the predicted value for the $ i $-th observation.
3) $ y_i $ is the actual value for the $ i $-th observation.
4) MAE provides a straightforward interpretation of the average error. Like RMSE, lower values indicate a better fit, and it has the same unit as the response variable.

The Coefficient of Determination ($ R^2 $) measures the proportion of variance in the dependent variable that is predictable from the independent variables. It essentially indicates how well the model fits the data. The $ R^2 $ value ranges from 0 to 1, with 1 indicating a perfect fit. The formula for $ R^2 $ is defined as:

\begin{equation}
R^2 = 1 - \frac{\sum_{i=1}^{n} (y_i - \hat{y}_i)^2}{\sum_{i=1}^{n} (y_i - \bar{y})^2}
\end{equation}

where:
1) $ n $ is the number of observations.
2) $ y_i $ is the actual value for the $ i $-th observation.
3) $ \hat{y}_i $ is the predicted value for the $ i $-th observation.
4) $ \bar{y} $ is the mean of the observed values.
5) An $ R^2 $ value close to 1 means that the model explains a large portion of the variance, whereas a value close to 0 indicates that the model explains very little variance.

\newpage
\section{Datasets comparison}

In this work, we compared the proposed method with the descriptor-based method \cite{orhan2021prediction}, MOFTransformer \cite{kang2023multi}, and PMTransformer \cite{park2023enhancing}. Although the eight datasets listed in Table 2 are derived from the study by \citet{orhan2021prediction}, their sizes differ slightly. The datasets used in this work were directly generated from the source repository (https://github.com/ibarisorhan/MOF-O2N2/tree/main/mofScripts) established by \citet{orhan2021prediction}. While the datasets used in MOFTransformer \cite{kang2023multi} and PMTransformer \cite{park2023enhancing} were also obtained from the same source, their exact data details were not explicitly provided. To ensure a clear and transparent comparison, we compiled the data information for all methods compared in this study, as summarized in Table~\ref{stbl:datadifference}.

\begin{table}[ht]
  \centering
  \caption{Comparison of datasets used in the published works across various MOF datasets.}
  \label{stbl:datadifference}
  \resizebox{1\textwidth}{!}{
  \begin{tabular}{ccccc}
  \hline
  \textbf{Datasets} & \textbf{CSTL} & \textbf{Descriptor-based}\cite{orhan2021prediction} & \textbf{MOFTransformer}\cite{kang2023multi} & \textbf{PMTransformer}\cite{park2023enhancing} \\
  \hline
  Henry's constant N\textsubscript{2}                                                 & 4744  & 4755 &      &      \\
  Henry's constant O\textsubscript{2}                                                 & 5036  & 5045 &      &      \\
  N\textsubscript{2} uptake (mol/kg)                                                  & 5132  & 5158 & 5286 & 5286 \\
  O\textsubscript{2} uptake (mol/kg)                                                  & 5241  & 5259 & 5286 & 5286 \\
  Self-diffusion of N\textsubscript{2} at 1 bar (cm\textsuperscript{2}/s)             & 5056  & 5079 & 5286 & 5286 \\
  Self-diffusion of N\textsubscript{2} at infinite dilution (cm\textsuperscript{2}/s) & 5192  & 5202 &      &      \\
  Self-diffusion of O\textsubscript{2} at 1 bar (cm\textsuperscript{2}/s)             & 5223  & 5247 & 5286 & 5286 \\
  Self-diffusion of O\textsubscript{2} at infinite dilution (cm\textsuperscript{2}/s) & 5097  & 5115 &      &      \\
  \hline
  \end{tabular}}
\end{table}

\section{Model Repeatability}\label{ssection:Repeatability}

\begin{figure}[!ht]
    \centering
    \includegraphics[width=16cm]{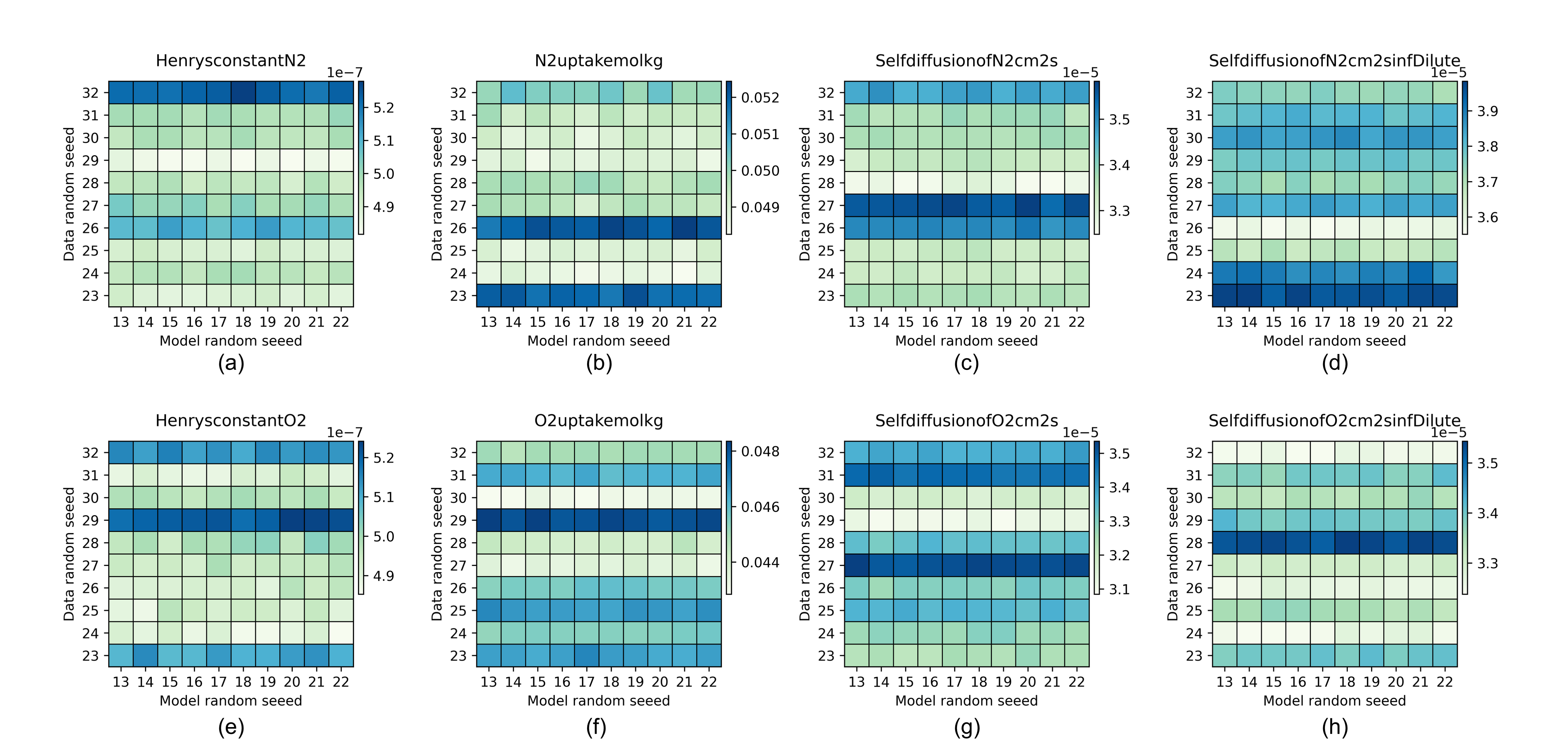}
    \caption{Heatmap of MAE values for predictive models across eight datasets related to O$_2$/N$_2$ selectivity properties in MOF materials. Panels (a)-(h) represent the MAE results for properties including Henry's constant for N$_2$ (a) and O$_2$ (e), N$_2$/O$_2$ uptake (mol/kg) for N$_2$ (b) and O$_2$ (f), self-diffusivity at 1 bar (cm$^2$/s) for N$_2$ (c) and O$_2$ (g), and self-diffusivity at infinite dilution (cm$^2$/s) for N$_2$ (d) and O$_2$ (h). Each dataset was randomly split 10 times with seeds ranging from 23 to 32, reserving 80\% for training and 10\% for testing. For each split, 10 separate models were trained with random seeds from 13 to 22, resulting in a total of 100 models per dataset. The heatmap color bar illustrates the MAE values for these 100 models, providing insight into prediction variability across different datasets and modeling scenarios.}
    \label{sfigure:trainmae}
  \end{figure}

\begin{figure}[!ht]
    \centering
    \includegraphics[width=16cm]{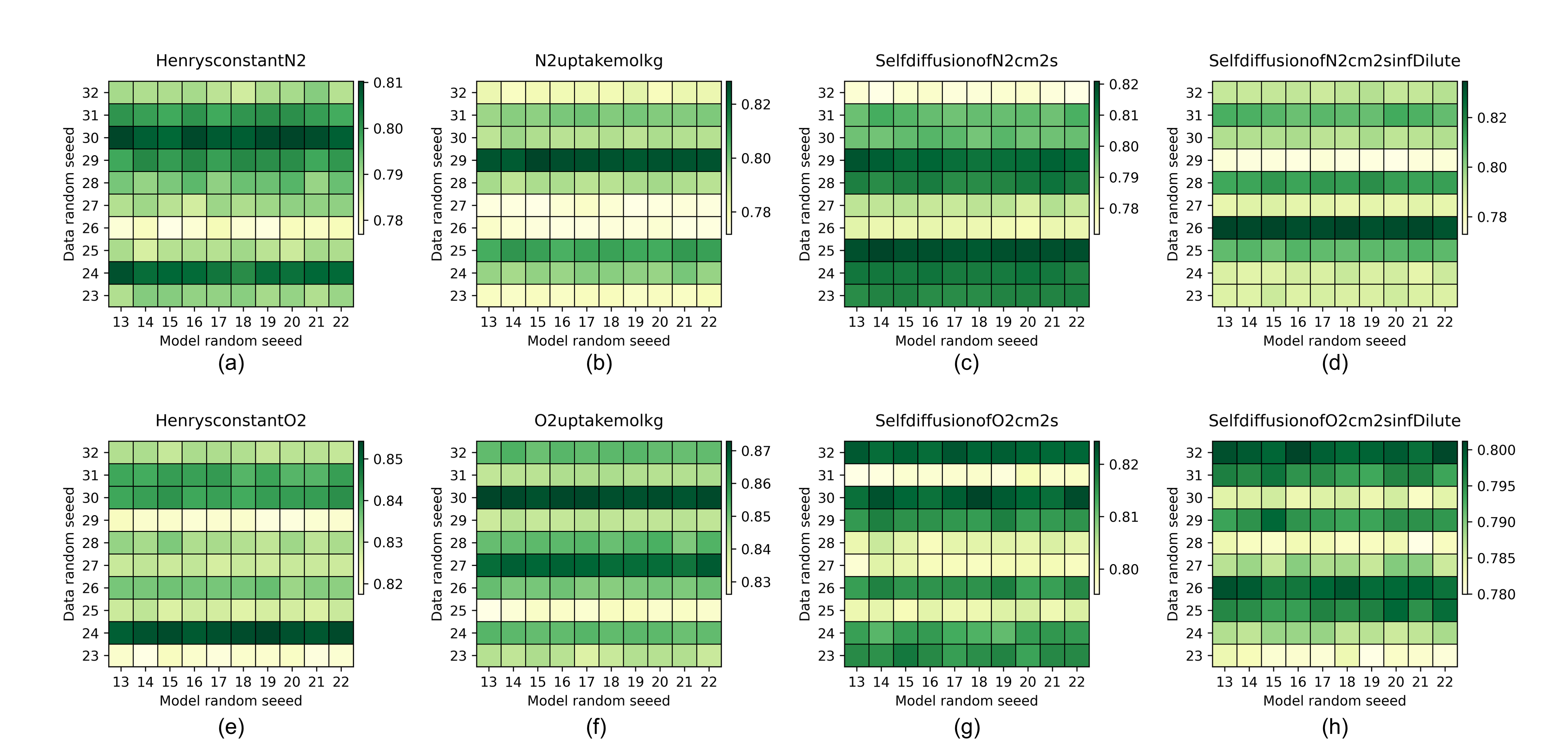}
    \caption{Heatmap of r$^2$ values for predictive models across eight datasets related to O$_2$/N$_2$ selectivity properties in MOF materials. Panels (a)-(h) represent the MAE results for properties including Henry's constant for N$_2$ (a) and O$_2$ (e), N$_2$/O$_2$ uptake (mol/kg) for N$_2$ (b) and O$_2$ (f), self-diffusivity at 1 bar (cm$^2$/s) for N$_2$ (c) and O$_2$ (g), and self-diffusivity at infinite dilution (cm$^2$/s) for N$_2$ (d) and O$_2$ (h). Each dataset was randomly split 10 times with seeds ranging from 23 to 32, reserving 80\% for training and 10\% for testing. For each split, 10 separate models were trained with random seeds from 13 to 22, resulting in a total of 100 models per dataset. The heatmap color bar illustrates the r$^2$ values for these 100 models.}
    \label{sfigure:trainr2}
  \end{figure}

\begin{figure}[!ht]
    \centering
    \includegraphics[width=16cm]{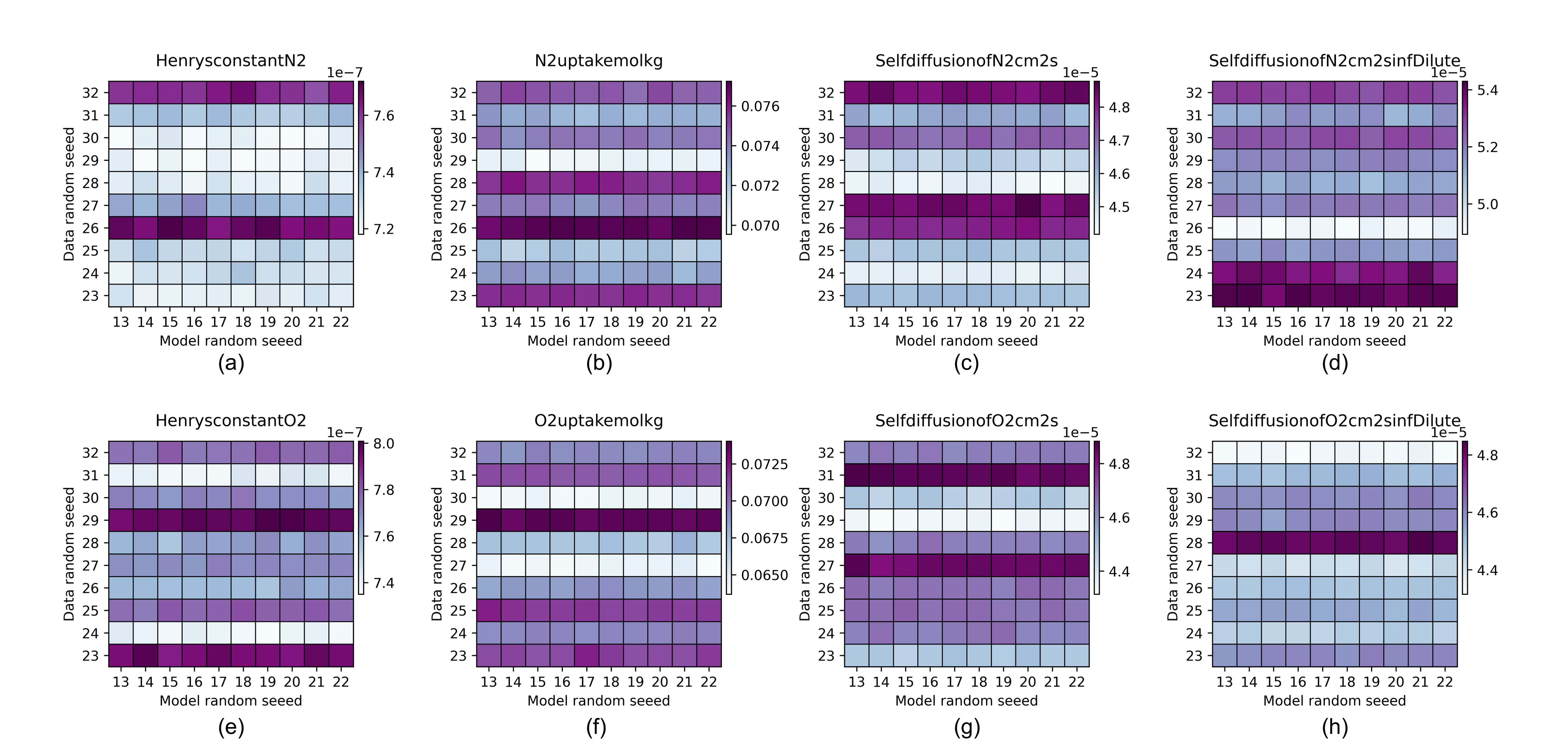}
    \caption{Heatmap of RMSE values for predictive models across eight datasets related to O$_2$/N$_2$ selectivity properties in MOF materials. Panels (a)-(h) represent the MAE results for properties including Henry's constant for N$_2$ (a) and O$_2$ (e), N$_2$/O$_2$ uptake (mol/kg) for N$_2$ (b) and O$_2$ (f), self-diffusivity at 1 bar (cm$^2$/s) for N$_2$ (c) and O$_2$ (g), and self-diffusivity at infinite dilution (cm$^2$/s) for N$_2$ (d) and O$_2$ (h). Each dataset was randomly split 10 times with seeds ranging from 23 to 32, reserving 80\% for training and 10\% for testing. For each split, 10 separate models were trained with random seeds from 13 to 22, resulting in a total of 100 models per dataset. The heatmap color bar illustrates the RMSE values for these 100 models.}
    \label{sfigure:trainrmse}
  \end{figure}

To ensure robust evaluation, we repeated the random data split 10 times, and for each split, 10 models were trained with different random seeds, resulting in a total of 100 models per dataset. The performance metrics, including RMSE, MAE, and $r^2$ correlation, were averaged over these 100 models and reported as the final results. This approach of using a single set of hyperparameters and a consistent evaluation protocol highlights the robustness of the predictive model, making the results reliable and comparable to existing methods in the literature. Specifically, the Gradient Boosting Tree (GBT) model was constructed to perform regression analysis using the proposed category-specific topological learning (CSTL) embedding as input features. We implemented the gradient boosting regressor from Scikit-learn \cite{pedregosa2011scikit}, optimizing the squared error loss function. The model parameters were set as follows: \text{max\_depth=7}, \text{max\_features=`sqrt'}, \text{min\_samples\_leaf=1}, \text{min\_samples\_split=2}, \text{n\_estimators=10,000}, and \text{subsample=0.5}. The heatmap of MAE/r$^2$/RMSE values for 100 predictive models across eight datasets are shown in Figure~\ref{sfigure:trainmae}, Figure~\ref{sfigure:trainr2}, and Figure~\ref{sfigure:trainrmse}.

\newpage

\section{Topological objects}\label{ssection:topo_represent}

\paragraph{Graph.}
The graph is a key structure for illustrating relationships among different entities, representing one of the most prevalent data forms. It is composed of nodes (or vertices) and edges, which establish the connections between these nodes. Graphs can be enhanced in several ways, such as by adding directionality to create directed graphs (digraph), assigning weights for weighted graphs, or incorporating geometric properties in geometric graphs. These enhanced graphs are excellent for representing relationships and attributes in various scenarios. Formally, a graph is defined as a pair $(V, E)$, where $V$ represents a set of vertices and $E$, a subset of $V \times V$, signifies the set of edges. Vertices and edges are the core components of a graph. Tools like adjacency matrices, degree matrices, and Laplacian matrices are utilized to describe the interactions between vertices and edges. These matrices are pivotal in graph theory and network analysis, capturing the graph's underlying topological structure. Although graphs are inherently one-dimensional, methods from simplicial complexes are sometimes used to express the graph's higher-dimensional aspects. 

\paragraph{Simplicial complex.}
A simplicial complex is a type of topological space constructed from basic units known as simplices. A simplex extends the notion of a triangle or tetrahedron to any number of dimensions. For a set of vertices $V$, a $k$-simplex $\sigma_k$ is typically represented by a subset of $V$ containing $k+1$ elements, and is expressed as $\sigma = \langle v_0, v_1, \ldots, v_k \rangle$. Any subset of $\sigma_{k-1}$ is considered a face of $\sigma_k$.

A \emph{simplicial complex}, denoted as $K$, based on a vertex set $V$, is defined by a group of simplices that meet two criteria: (1) If a simplex $\sigma$ is part of $K$, then all of its faces, including individual vertices, are also included in $K$; (2) The intersection of any two simplices within $K$ is either empty or a face (subset) common to both simplices. From these characteristics, it's evident that a graph can be interpreted as a 1-dimensional simplicial complex, where its simplices consist of vertices (0-simplices) and edges (1-simplices).

In a $k$-simplex, the boundary is the set of its $(k-1)$-dimensional faces. The \emph{boundary operator}, symbolized as $\partial_{k}$, operates on a $k$-simplex $\langle v_0, v_1, \ldots, v_k\rangle$ in the following mathematical form:
\begin{equation}
    \partial_{k}  \langle v_0, v_1, \dots, v_k \rangle= \sum_{i=0}^{k} (-1)^i \langle v_0, \dots, \widehat{v_i}, \dots, v_k\rangle ,
\end{equation}
where $\widehat{v_i}$ indicates the exclusion of the vertex $v_i$. A chain complex is a series of Abelian groups (or modules), interconnected by boundary operators. Suppose $G$ is an Abelian group. The $k$-th group in the chain complex, denoted as $C_{k}(K;G)$, comprises formal sums of $k$-simplices. The boundary operator $\partial_k : C_{k}(K;G) \to C_{k-1}(K;G)$ maps a $k$-simplex to its $(k-1)$-dimensional boundary. The sequence of the chain complex can be represented as:
\begin{equation}
    \cdots  \xrightarrow{\partial_{k+1}} C_{k} (K;G)\xrightarrow{\partial_k} C_{k-1}(K;G) \xrightarrow{\partial_{k-1}} \cdots \xrightarrow{\partial_2} C_{1} (K;G) \xrightarrow{\partial_1} C_{0}
    (K;G).
  \end{equation}
A critical characteristic of the boundary operator is that the composition of two consecutive boundary operators equals zero, i.e., $\partial_{k-1} \circ \partial_k = 0$. This implies that the boundary of a boundary is always null, carrying significant topological implications. The structure of the chain complex provides a systematic way to analyze how boundaries integrate with each other. Beyond the simplicial complex, other topological objects—such as the clique complex, cell complex, cellular sheaf \cite{wei2023persistent}, hypergraph, neighborhood complex \cite{liu2023neighborhood, liu2022neighborhood}, Hom complex, knot, link, and tangle \cite{kozlov2007combinatorial, shen2024evolutionary}—can be further explored in the analysis of the given data.


\newpage
\section{Supplementary tables}\label{ssection:supplementary_tables}

  In the following section, we provide supplementary tables that offer additional data and insights pertinent to our study. Readers are encouraged to refer to these tables for a more detailed exploration of the topics covered in the main text.

 \begin{table}[!ht]
  \centering
  \caption{Comparison of CSTL models with different training-test splits.}
  \label{stbl:comparison}
  \resizebox{1\textwidth}{!}{
  \begin{tabular}{lcccccc}
  \hline
  \multirow{2}{*}{Datasets} & \multicolumn{3}{c}{CSTL(80\% training, 10\% test)} & \multicolumn{3}{c}{CSTL(80\% training, 20\% test)} \\ \cline{2-7}
   & $r^2$ & mae & rmse & $r^2$ & mae & rmse \\ \hline
  Henry's constant N$_2$                              & 0.80 & 4.90E-07 & 7.25E-07 & 0.79 & 4.98E-07 & 7.36E-07 \\
  Henry's constant O$_2$                               & 0.83 & 4.98E-07 & 7.63E-07 & 0.83 & 5.00E-07 & 7.69E-07 \\
  N2 uptake (mol/kg)                                & 0.79 & 4.98E-02 & 7.37E-02 & 0.79 & 4.98E-02 & 7.39E-02 \\
  O2 uptake (mol/kg)                                & 0.85 & 4.50E-02 & 6.82E-02 & 0.85 & 4.54E-02 & 6.90E-02 \\
  Self-diffusion of N$_2$ at 1 bar (cm2/s)             & 0.80 & 3.40E-05 & 4.69E-05 & 0.80 & 3.39E-05 & 4.64E-05 \\
  Self-diffusion of N$_2$ at infinite dilution (cm2/s) & 0.80 & 3.75E-05 & 5.15E-05 & 0.80 & 3.79E-05 & 5.21E-05 \\
  Self-diffusion of O$_2$ at 1 bar (cm2/s)             & 0.82 & 3.21E-05 & 4.45E-05 & 0.81 & 3.32E-05 & 4.62E-05 \\
  Self-diffusion of O$_2$ at infinite dilution (cm2/s) & 0.79 & 3.34E-05 & 4.53E-05 & 0.79 & 3.35E-05 & 4.54E-05 \\
  \hline
  \end{tabular}}
  \end{table}

\newpage

\section{Supplementary figures}\label{ssection:supplementary_figures}
  In this section, we present a series of supplementary figures that further elucidate and complement the findings discussed in the main text. Readers are encouraged to consult these figures for a richer understanding and visual representation of the concepts and results introduced in the main manuscript.

  \begin{figure}[!ht]
    \centering
    \includegraphics[width=16cm]{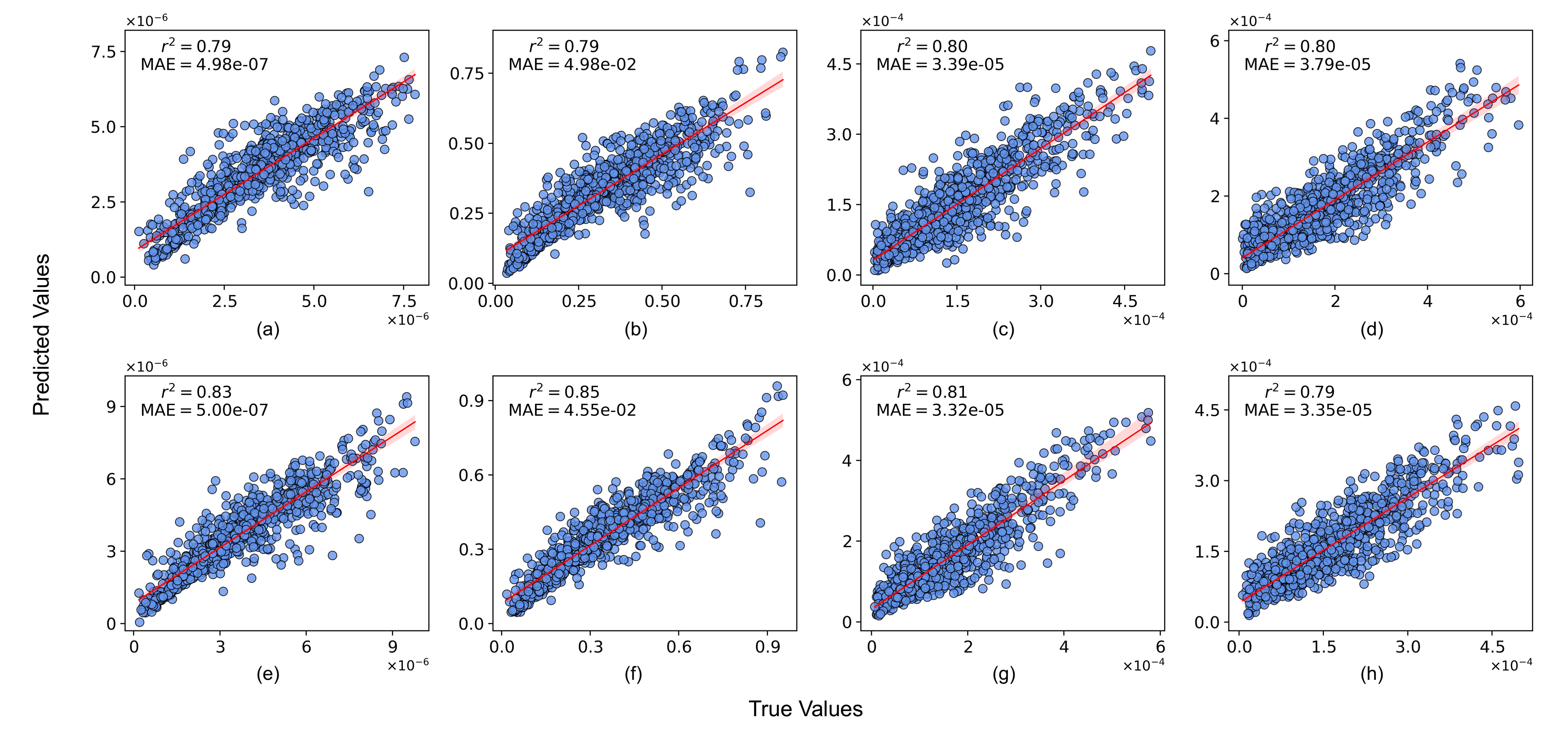}
    \caption{Comparison between predicted and true values for eight datasets on O$_2$/N$_2$ selectivity properties in MOF materials. Panels {\bf a}-{\bf h} show prediction performance for different properties: Henry's constant for N$_2$/O$_2$ (a, e), N$_2$/O$_2$ uptake (mol/kg) (b, f), self-diffusivity of N$_2$/O$_2$ at 1 bar (cm$^2$/s) (c, g), and self-diffusivity of N$_2$/O$_2$ at infinite dilution (cm$^2$/s) (d, h). Each panel displays the R$^2$ and the MAE in the upper left corner. Each dataset was randomly split, with 80\% used for training and rest 20\% for testing.
    }
    \label{sfigure:preditions}
  \end{figure}

  \newpage

  \newpage

  \newpage

  \newpage
  \begin{figure}[!ht]
    \centering
    \includegraphics[width=13cm]{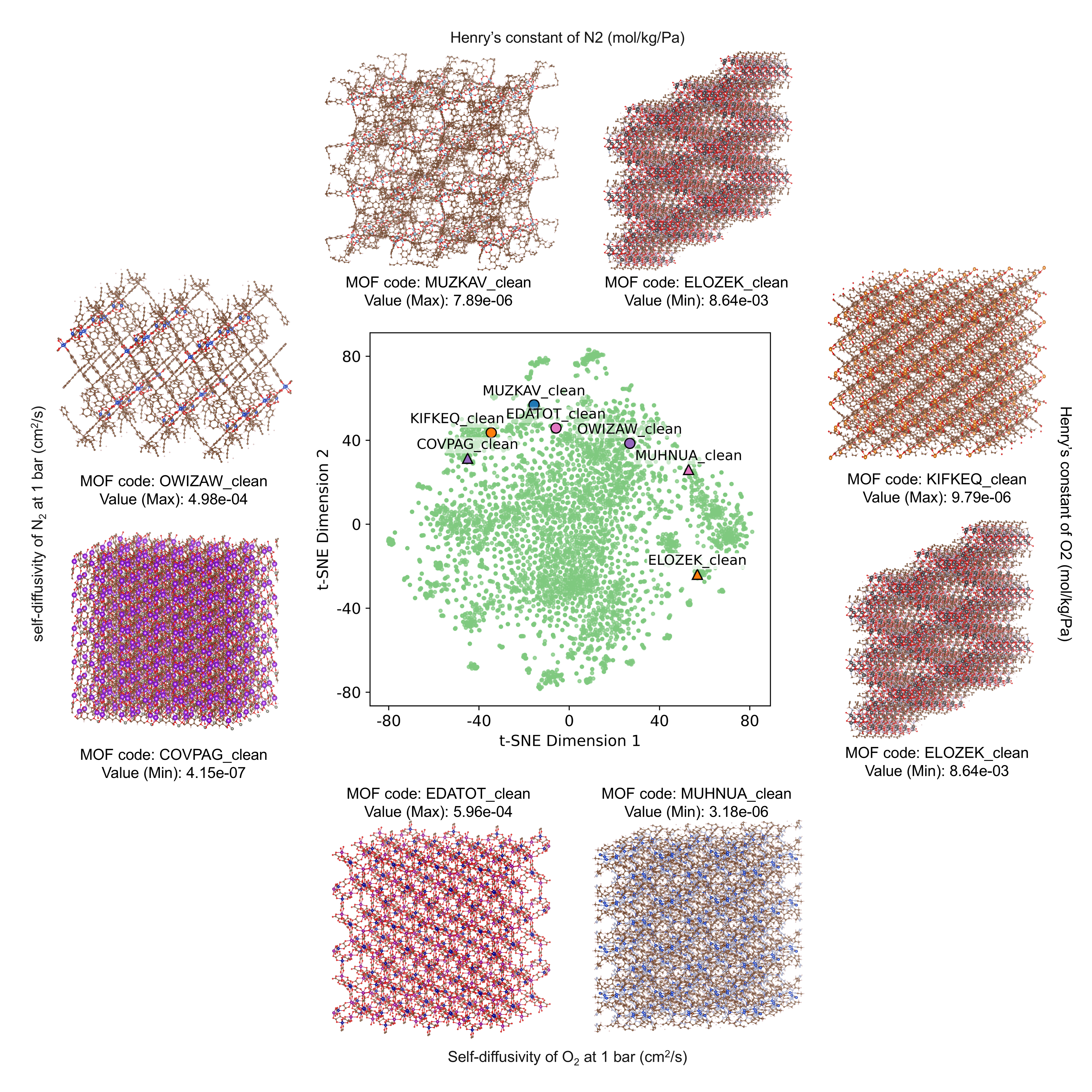}
    \caption{t-SNE feature reduction for category-specific topological features of MOF materials, where each green point represents a distinct MOF material. Highlighted circles and triangles indicate materials with maximum and minimum values, respectively, for four key properties: Henry's constant for N$_2$, Henry's constant for O$_2$, self-diffusivity of N$_2$ at 1 bar (cm$^2$/s), and self-diffusivity of O$_2$ at 1 bar (cm$^2$/s). 3D structures of the materials with minimum and maximum values for each property are shown around the t-SNE plot.}
    \label{sfigure:embedding2}  
  \end{figure}

\newpage

\end{document}